# Odor response features of projection neurons and local interneurons in the honeybee antennal lobe.

*Anneke Meyer (1,2), Giovanni Galizia (2), and Martin Paul Nawrot (1,3)*

1 Theoretical Neuroscience, Institute of Biology, Freie Universität Berlin, 14195 Berlin, Germany
2 Department of Biology, University of Konstanz, Konstanz, Germany
3 Bernstein Center for Computational Neuroscience (BCCN) Berlin, 10115 Berlin, Germany

## Abstract

Local computation in microcircuits is an essential feature of distributed information processing in vertebrate and invertebrate brains. The insect antennal lobe represents a spatially confined local network that processes high-dimensional and redundant peripheral input to compute an efficient odor code. Social insects can rely on a particularly rich olfactory receptor repertoire and they exhibit complex odor-guided behaviors. This corresponds with a high anatomical complexity of their AL network. In the honeybee, a large number of glomeruli that receive sensory input are interconnected by a dense network of local interneurons (LNs). Uniglomerular projection neurons (PNs) integrate sensory and recurrent network input into an efficient spatio-temporal odor code. To investigate the specific computational roles of LNs and PNs we measured eleven features of sub- and suprathreshold single cell responses to *in vivo* odor stimulation. Using a semi-supervised cluster analysis we identified a combination of five characteristic features that enabled the accurate separation of morphologically identified LNs and PNs. The two clusters differed significantly in all five features. In the absence of stimulation PNs showed a higher subthreshold activation, assumed higher peak response rates and more regular spiking pattern. LNs reacted considerably faster to the onset of a stimulus and their responses were more reliable across stimulus repetitions. We discuss possible mechanisms that can explain our results, and we interpret cell-type specific characteristics with respect to their functional relevance.

## Introduction

Sensory computation in the nervous systems of both, invertebrates and vertebrates, is organized in local networks containing microcircuits that integrate local feed-forward and recurrent connections and constitute functional subunits of the global sensory network. Understanding the computational principles of these microcircuits is a key to a deeper understanding of sensory processing and perception (Chou et al., 2010; Shepherd, 2010). As a common principle microcircuits are built from synapses between two general types of neurons, local interneurons (LNs) and projection neurons (PNs). Neurites of LNs are spatially confined to a local brain structure while PNs connect between brain structures. Both, network connectivity and the individual morphological and physiological properties of LNs and PNs define the function and reflect the specific processing demands of a particular sensory system.

Primary olfactory centers, the vertebrate olfactory bulb and the analogue invertebrate antennal lobe (AL), perform complex local computations (Olsen and Wilson, 2008a; Sachse et al., 2006; Strowbridge, 2010) that reflect the high dimensionality of the chemical olfactory space (Guerrieri et al., 2005; Haddad et al., 2008; Schmuker and Schneider, 2007; Wilson and Mainen, 2006) as well as the complex temporal dynamics of natural odor stimuli (Meyer and Galizia, 2012; Nagel and Wilson, 2011; Riffell etal., 2009; Stopfer, Jayaraman, and Laurent, 2003). At the heart of these computations are the glomeruli, prominent examples of sensory microcircuits. They form spherical structures of high synaptic density in which peripheral input from olfactory sensory neurons (OSNs) converges onto LNs and PNs. In the present study, we explore differences in *in vivo* odor response properties between LNs and PNs in the primary olfactory center of the honeybee.

In the invertebrate, structural complexity of the AL correlates with the complexity of odor-guided behavior in individual species. Anatomical complexity is particularly pronounced in social insects such as bees and ants (Galizia and Rössler, 2010; Kelber et al., 2010; Martin et al., 2011; Zube and Rössler, 2008). The interneuron network interconnects different glomeruli and thus plays an essential role in olfactory information processing (Abraham et al., 2004; Chou et al., 2010; Galizia and Kimmerle, 2004; Kazama and Wilson, 2009; Krofczik et al., 2009; Meyer and Galizia, 2012; Olsen and Wilson, 2008b; Sachse and Galizia, 2002). The number of LNs largely determines the degree of network connectivity and hence its computational capacity. In the honeybee approximately 4. 000 LNs outnumber PNs almost fivefold, providing for an exceptionally dense interneuron network. Despite the obvious importance of the interneuron network we know surprisingly little about its detailed involvement in sensory computation (Galizia, 2008; Nawrot, 2012).

For our analyses we combined independently obtained data sets from *in vivo* intracellular recordings of neurons in the honeybee AL. A subset of cells could be identified unambiguously as either LN or PN. This allowed us to devise a semi-supervised clustering method for separating the two cell classes on the basis of their electrophysiological responses. Our results provide characteristic differences in odor response features for PNs and LNs indicating their differential role in computing the spatio-temporal odor code that is conveyed to central brain structures.

## Materials & Methods

*Data sets*

Analysis of odor evoked activity patterns was performed on intracellular recordings from 80 AL neurons. The data pool comprised three independently obtained and published data sets as well as one set of unpublished data (n = 10 cells). The same recording technique was used in all cases, but different sets of primary odorants and odor mixtures were tested. The minimum stimulus duration within all experiments was 800ms. For details of data acquisition and tested odor sets refer to the original works by Meyer and Galizia (2011), Krofczik et al. (2008), and Galizia and Kimmerle (2004). Based on morphological data from staining or in some PN recordings the position of the recording site (Krofczik et al. 2008), a subset of cells could be identified as PNs (n = 32) or LNs (n = 10).

*Data preprocessing*

Potent stimuli, i. e. stimuli that evoked responses, were identified for each individual cell by visual inspection. Spike-times were detected using Spike2 (Cambridge Electronic Design, UK) or custom written routines in R (http://www. R-project. org) based on the open source packages SpikeOMatic (Pouzat et al., 2004) and STAR (Pippow et al., 2009). To describe sub-threshold characteristics we removedall action potentials from the raw signal using a custom written routine in MatLab (7. 10. 0, The Mathworks Inc., MA).

*Determination of optimal feature set*

Neural responses were analyzed in the response window $W_{resp}$= *[0ms,800ms]* following stimulus onset (t=0ms) and baseline activity was analyzed in the baseline window $W_{base}$ = *[-500ms,0ms]* immediately preceding stimulus onset (Fig 1A). We defined a total of 11 electrophysiological features that describe different properties of neural response activity: the trial-averaged baseline firing rate, change in firing rate (ΔR), response latency (L), variance of response latency between odors, coefficient of variation (CV2), Fano factor (FF), trial-averaged spontaneous power ($P_{base}$), stimulus related power, integrated membrane potential, maximal hyper- and depolarization of the membrane potential integrated over the area of deflection. Our goal was to find an optimal subset of features that allows separating the two morphological classes of LNs and PNs. This combination of descriptors was found by testing cell type classification forall possible feature combinations in a repeated semi-supervised clustering procedure. The core routine of the semi-supervised method was identical with the one detailed below for the final clustering result. In brief, the selected combination of descriptors was submitted to PCA. The number of informative PCs was selected from the scree-plots. Clustering was performed on the determined number of PCs and the number of clusters was fixed to two. We calculated the separation quality of identified neurons in the two clusters using Matthew's Correlation Coefficient (Matthews, 1975). By this procedure we identified a subset of five relevant features that yielded the best separation of PNs and LNs. These were rate change, latency, CV2, Fano-Factor, and spontaneous power. Their computation is detailed below. For analysis and visualization of the data we used Matlab.

*Definition of response features*

*ΔR:* Deflection from the baseline firing rate immediately following stimulus application is the most common definition of evoked spiking activity. Rate increase (decrease) is a measure for excitation (inhibition). The time-resolved firing rate profile was estimated based ontrial-aligned and trial-averaged spike-trainsfollowing the method described in Meier et al. (2008). In brief: First, the derivative of each single trial spike-train of a given cell under stimulation of one odor was estimated by convolving the spike train with an asymmetric Savitzky-Golay filter (Savitzky and Golay, 1964) (polynomial order 2, width 300ms, Welch-windowed). Second, all single trial derivatives were optimally aligned by maximizing their average pair-wise cross correlation (Nawrot et al., 2003) (Fig 1C). Third, the newly aligned spike-trains were merged. Fourth, the alignment procedure was repeated for the merged spike-trains of different odors. To estimate the trial-averaged time-resolved rate profile the pooled across-odor spike-train was convolved with an asymmetric alpha kernel $k(t) = t * \exp(-t/\tau)$ (Parzen, 1962)(Fig 1D). Optimal kernel width $\tau$ was estimated on the basis of the empirical data by application of a heuristic method detailed in (Nawrot et al., 1999). *ΔR* was then defined as the difference between the highest value of peak firing rateand the minimum rate value encountered in any of the trials, irrespective of the odor. Thus *ΔR* estimates the maximal modulation depth of firing rate across time and odors.

*L:* describes the positive time interval between stimulus onset and onset of neural response. Trial-averaged absolute latency and relative trial-to-trial latencies were estimated with one of three methods based on the cell's firing pattern. 1) Latencies with excitatory responses were estimated based on the derivative of the trial-aligned firing rate (Meier et al., 2008; Krofczik et al., 2008). The trial alignment procedure was conducted as described above. By convolution of the summed across-odor spike-train with the same asymmetric Savitsky-Golay filter that was used for the alignment procedure, an estimate about the derivative of the cell's average firing rate was obtained. The cell specific absolute latency was defined as the time point of the first maximum encountered in the derivative (Fig 1C). 2) Latencies of inhibitory responses were estimated identically but using an inverted Savitsky-Golay filter to detect the maximum of the negative slope. 3) Latencies of cells that had very low spontaneous activity and which responded to stimulation with a membrane depolarization accompanied by one single or very few spikes were estimated based on the pooled original spike-trains and not aligned. Spikes denoting a response were generally well timed. An additional alignment usually introduced faulty shifts as a consequence of the generally low spiking activity. The response latency was thus defined as the peak-time of the rate, which in these conditions essentially resembled the first spike latency. Rate was estimated as detailed above.
To normalize absolute latencies for differences in odor delivery times in the different data sets which arise from differences in the experimental setup we preceded as follows: At any one time we subtracted the shortest latency within each individual data set from all other latency estimates within the same data set. To avoid zero latency, we added the arbitrary duration of 6ms to the response latency of each cell.

*CV2*: indicates a neuron's spike-time irregularity from the inter-spike intervals (Nawrot, 2010) (Fig 1C). The *CV2* was introduced to quantify interval dispersion when firing rate is not constant but modulated (Holt et al., 1996; Ponce-Alvarez et al., 2010). It is defined locally as the variance of two consecutive ISIs divided by their mean. We first calculated the averaged CV2 for each single trial and then averaged over all trials, irrespective of stimulus type.

*FF:* is an established measure for spike count variability (Nawrot et al., 2008) and defined by the ratio of the across-trial variance and the trial-averaged spike countwithin $W_{resp}$. We computed the FF for each stimulus separately and subsequently averaged across odors.

*$P_{base}$:* of the membrane potential (after spike removal, Fig 1B) during the pre-stimulus interval $W_{base}$ quantifies the membrane potential fluctuations in the absence of a driving stimulus. It is computed within each trial as $P = 1/T \int_{T}^{0} |s(t)|^2 \, dt$ and subsequently averaged across trials.

*Cluster analysis.*

Collecting descriptive values to characterize evoked activity results in a multi-dimensional data space. Several descriptors derive in part from the same origin and may hence be correlated and carry partly redundant information. Principal Component Analysis (PCA) allows to reduce a set of possibly correlated variables into a smaller set of uncorrelated variables called Principal Components (PC) (Pearson, 1901) that still retain the major information content. Using PCA in the present dataset allowed reducing five descriptors to the first three PCs. These were sufficient to explain 75% of the underlying variance. Since the original variables differ in the scale on which observations was made, data was normalized using z-scores before it was subjected to the PCA algorithm. To explore possible grouping of neurons according to the PCs of their evoked activity characteristics, unsupervised clustering using Ward linkage with Euclidean distances was performed. The incremental method aims to reduce the variance within a cluster by merging data points into groups in a way that their combination gives the least possible increase in the within-group sum of squares (Ward, 1963). The sum of squares as the distance measure $d$ between two groups (r,s) is defined as:

$$d(r,s) = \sqrt{\frac{2 n_r n_s}{n_r + n_s}} \parallel \bar{x}_r - \bar{x}_s \parallel_2$$

Where $\parallel \parallel_2$ denotes the Euclidean distance, $\bar{x}r$ and $\bar{x}s$ are the centroids of clusters $r$ and $s$, and $n$ refers to the number of elements in each cluster. The algorithm was provided by the Matlab Statistics Toolbox. In order to test whether clustering performed on simplified PC inputyields information, which allows describing neuron differences in terms of direct measurable characteristics, we performed a rank sum tests on the features between the two clusters.

## Results

**Classification of PNs and LNs can be achieved based on an optimal set of electrophysiological response features.**

We initially defined eleven distinct measures of electrophysiology to describe the response properties of each of the 80 AL neurons in our data set (see Materials &Methods). To classify LNs and PNs (Fig 2A) we applied a semi-supervised clustering method based on all possible combinations of 11 features and evaluated the outcome using separation of morphologically identified LNs and PNs as a measure for model quality. By systematic variation of the feature set and the dimension of the principal component (PC) space we found that several subsets of our measures were sufficient to separate identified LNs and PNs significantly above chance level. We aimed at finding that constellation in which the best classification could be achieved based on a minimal set of input features. The most efficient solution allowed for a correct classification of 39 out of 42 identified neurons corresponding to a Mathew's correlation coefficient of 0.78. It is based on the first three PCs (75% explained variance) from a combination of five response features: change in firing rate from baseline (ΔR), response latency (L), CV2, Fano factor (FF), and spontaneous signal power ($P_{base}$). In an attempt to visualize functional stereotypy we arranged one randomly selected spike train from each neuron (Fig 2B) according to their relationship in the cluster tree (Fig 2C). Judging from this account it appears that neurons in the PN cluster display aphasic-tonic response with high rate changes. LN cluster neurons are, in comparison, marked by phasic but curbed responses. However, a clear-cut classification of neurons by eye seems impossible. To visualize spatial separation of the PN and LN dominated clusters we plotted all cells in the three dimensional PC space (Fig 2D). The two clusters largely separate from each other but do show an area of overlap, in which misclassification is more likely to appear. To further quantify cluster quality we compared the distribution of distances of individual elements to the cluster centers within and between the clusters (Fig 2E). Distances within each of the clusters are clearly shorter than between the clusters.

**LNs and PNs differ significantly in their odor response features.**

We could show that based on the PCs of five electrophysiological measures, neurons clustered in two groups, one of which is clearly dominated by PNs, the other by LNs. Next we asked if this clustering is reflected in significant differences in the input feature space, i. e. the actual odor response measures. Indeed, we found that the PN and the LN cluster differed significantly in each of these measures (Wilcoxon rank sum test, Table 1; Fig 3). Neurons in the PN cluster typically showed higher dynamic changes in firing rate when responding to a stimulus. This is in good accordance with the observed tendency for phasic-tonic response patterns (Fig 2B). The curbed responses of LNs typically follow stimulus onset with shorter response latencies than PNs. The difference in median latencies between LNs and PNs is considerable with 65ms. Interestingly, latencies in both clusters show a broad distribution across neurons. Particularly, response onsets in the subset of identified LNs varies between quartiles by about 200ms ($1^{st}$ quartile = 36ms, $3^{rd}$ quartile = 235ms). Response onsets in the subset of identified PNs is less variable with an inter-quartile distance of about 100ms ($1^{st}$ quartile = 74ms, $3^{rd}$ quartile 170ms). The higher CV2 for neurons allocated to the LN cluster illustrates that these cells are characterized by more irregular or burst-like spike responses, while cells of the PN cluster show more regular response trains. A higher Fano factor indicates responses from PN cluster neurons to be more variable across trials.

Differences in all five features between neurons in the LN and PN cluster transfer to the subset of morphologically identified neurons. This reassures that electrophysiological characteristics are truly stereotyped properties of LNs and PNs, respectively (Table1; Fig. 3). Change in response related firing rate and CV2 in particular are significantly different (p<0. 05) even for the small sample size of identified LNs (N = 10) and PNs (N = 32). For latency, Fano factor, and spontaneous signal power, differences in median for morphologically identified LNs and PNs are in accordance with the respective differences measured on the basis of the complete set of neurons.

## Discussion

Based on intracellular recordings from a mixed neuron population in the honeybee AL we explored characteristic differences between LNs and PNs. Electrophysiological measures are established means by which neurons are typified if morphological information is unavailable (Connors and Gutnick, 1990; Ascoli et al., 2008; Markram et al., 2004). Clustering analyses have been used repeatedly in vertebrates to typify neurons on the basis of morphological and electrophysiological features, and in order to characterize their specific functional properties within microcircuits (McCormick et al., 1985; Ruigrok et al., 2011; Suzuki & Bekkers, 2006, 2011; Wiegand et al., 2011). In our approach we clustered cells solely based on physiological response measures to separate two morphologically well described classes of LNs and PNs in the honeybee AL. Using the morphological class identity available for a subset of all cells allowed us to assess classification accuracy and to optimize the clustering approach with respect to the number of PCs, and the particular combination of features. We found a combination of five out of eleven odor response features to be indicative of the morphological cell type. How can we interpret these characteristic physiological differences in a functional context?

**PN properties are well suited to convey a combinatorial rate code.**
A considerable level of spontaneous activity and a strong and odor-specific modulation of the firing rate (Fig. 3) have been described as characteristic for honeybee PNs, but less typical for LNs in independent comparative studies (Abel et al., 2001; Müller et al., 2002; Sun et al., 1993). Pronounced baseline activity may arise from cell-intrinsic excitability or auto-rhythmic activity in the absence of input, or from ongoing network input (Llinas, 1988). Baseline activity in AL neurons was recently shown to depend on continuing OSN input even in the absence of overt stimuli and not on auto rhythm (Joseph et al., 2012). PNs form numerous synapses with both LNs and a large number of converging OSNs (Distler and Boeckh, 1997; Galizia, 2008). During odor stimulation PNs are the object of strong afferent OSN input and recurrent network input. According to our analysis PNs expressed prominent rate modulations, with typical peak rates in the order of 50-100Hz. The PN population is thus well suited to project a spatio-temporal rate code to the higher brain centers. Evidence for the existence and behavioral relevance of a combinatorial odor rate code in the PN ensemble has been provided by a number of recent extracellular recordings (e. g. Brill et al., 2012; Strube-Bloss et al., 2012; Strube-Bloss et al., 2011; Müller et al., 2002).

**Irregular spiking and short latencies reflect the modulatory function of LNs.**
The local interneuron network provides the substrate for mediating a non-linear transformation between AL input and output in flies and bees (Bhandawat et al., 2007; Ng et al., 2002; Olsen and Wilson, 2008; Sachse et al., 2006; Meyer and Galizia, 2011; Schmuker 2012). A prerequisite is the widely ramified LN morphology that interconnects many different glomeruli, integrating information from different genetic receptor types. The high CV2 of LNs (Fig. 3) likely is a physiological reflection of this intertwined connectivity. Spike time irregularity arises from two events: when inhibitory input counteracts excitatory input (Vreeswijk and Sompolinsky, 1996; Shadlen and Newsome, 1998; Stevens and Zador, 1998; Nawrot et al., 2008), or when the excitatory inputs arrive in an irregular fashion, e.g. through integration of inputs with different spike train statistics (Renart et al., 2010 ; Farkhooi et al., 2011), and output irregularity is particularly high when both conditions apply (Bures, 2012). Irregular LN output is likely a consequence of heterogeneous input from both, excitatory (OSNs and PNs) and inhibitory (LNs) sources (Malun, 1991; Galizia and Rybak, 2010). In addition, the superposition of inputs from several co-activated glomeruli makes excitatory input irregular.

A striking result of our analysis is the faster response time of LNs with a median response latency of only ~60ms compared to ~120ms for PNs (Table 1). Fast LN responses coincide with the previous observation of an equally fast reduction of the membrane potential in single PNs (Krofczik, Menzel & Nawrot, 2008) and indicate that LNs can efficiently modulate PN output through fast lateral inhibition. The distribution of individual latencies is rather broad in both neuron populations (Fig. 3). Single PNs can respond much faster than the population average. This observation is interesting in light of the recent findings by Strube-Bloss, Herrera-Valdez & Smith (2012) that AL neurons responded, on average, later to odor stimulation than mushroom body (MB) output neurons, which are situated two synapses downstream of

PNs. Meyer & Galizia (2011) tested responses of AL neurons to a mixture with two components. They found elemental neurons that showed fast responses dominated by and temporally locked to the dominant mixture component. In contrast, configural neurons that represented the novel mixture quality showed longer response latencies. Together this may indicate that a fast PN population carries an initial rapid odor code. Recurrent projections from the MB to the AL modulate a secondary delayed odor code. It has been suggested that the early phasic stimulus response component establishes a latency code of odor identity in the insect (Krofczik et al., 2008; Kuebler et al., 2011; Brill et al., 2012), which might be required for rapid behavioral action. A late and persistent odor code might support the refined percept of the stimulus environment, e. g. mixture composition and concentration of individual elements (Fernandez et al., 2009; Strube-Bloss, Herrera-Valdez & Smith, 2012), and it might underlie the formation of associations.

**Properties of AL neurons differ between species.**
Throughout species the AL is organized in a glomerular fashion and built from the same elements: OSNs, PNs and LNs. However, numbers and wiring of these constituents differs vastly between species. As a consequence PNs and LNs may well exhibit different physiological properties in different species. The AL of the Tobacco Hornworm *Manduca sexta* has regular spiking LNs and shows irregular, burst like activity in PNs (Lei et al., 2011), opposite to our findings. In *Drosophila*, populations of both regular and irregular spiking LNs have been described (Chou et al., 2011; Seki et al., 2010). In the cockroach neurons were identified, which produce sodium spikes (Husch et al., 2009). In the locust, only non-spiking interneurons were found so far (Laurent, 1993). An explanation for these physiological variations might be found in the species specific architecture. About 160 glomeruli in the honeybee AL are connected with ~4000 LNs (Withöft, 1967) but give output via only ~800-900 PNs (Rybak 2010). Honeybee LNs innervate subareas of glomeruli in which OSN input is concentrated as well as subareas in which PN neurites dominate (Fonta, 1993), and LNs are likely to form inter- as well as intra glomerular connections (Meyer and Galizia 2011). In other prominent insect models for olfaction LNs are less numerous than PNs and the overall degree of connectivity is much smaller (*Drosophila*: <50 glomeruli (Stocker, 1994), 150-200 PNs (Stocker, 1997), 100 LNs (Ng et al., 2007); locust: 830 PNs (Leitch and Laurent, 1996), 300 LNs (Anton and Homberg, 1999); moth: ~60 glomeruli (Sanes and Hildebrand, 1976b), 740 PNs, 360 LNs (Homberg, 1988)). Naturally, these differences in architecture are not only reflected in physiological properties of single neurons but impact the entire network function at the level of odor and odor mixture encoding, which seems necessary for the species-specific adaption to environmental constraints (Martin et al., 2011).

**The diversity of AL neurons within species**
LNs and PNs establish two anatomically and morphologically well-defined classes of AL neurons. However, both display considerable within-class diversity. In some species PNs subdivide in morphological subgroups (Galizia and Rössler, 2010). In most hymenoptera, including the honeybee, PNs subdivide into three morphological families (Rösler and Zube, 2011). LNs can show various different morphologies within a species (Chou et al. 2010; Christensen et al. 1993; Dacks et al. 2010; Flanagan and Mercer 1989; Fonta et al. 1993; Seki and Kanzaki 2008; Seki et al. 2010; Stocker et al. 1990). In the honeybee so-called homogeneous and heterogeneous LNs represent two major subgroups. However, even morphologically similar LNs may be further differentiated according to, for instance, their histochemistry (Dacks et al., 2010; Kreissl et al., 2010; Nässel and Homberg, 2006; Schäfer and Bicker, 1986; Chou et al., 2010; Ng et al., 2007). The existence of different families is supported by the diversity of LN physiology (Chou et al., 2011; Seki et al., 2011; Husch et al., 2009; Sachse et al., 2003; Meyer and Galizia, 2011) that finds expression in the variances of individual response properties within the LN group of our data set (Fig. 3) and explains why we could not achieve 100% accuracy of classification (Fig. 2). In future work it will be desirable to extend the present approach to extract communal features of known subgroups such as homo and hetero LNs, or PN families. While our analysis still provides a limited picture of honeybee LN- and PN-physiology, it provides for the first time systematic differences of their response physiology. Such detailed knowledge is essential to foster realistic models of neural computation that can explain the complex spatial and temporal processing of peripheral olfactory information in the primary olfactory center.

## Acknowledgments

We are grateful to Randolf Menzel, Sabine Krofczik and Bernd Kimmerle for providing us with their data sets for re-analysis in the present manuscript. We thank Jürgen Rybak for his assistance with the morphological data, and Michael Schmuker and Jan Sölter for methodological consultancy. Generous funding was received from the German Federal Ministry of Education and Research (BMBF) within the project *Bernstein Focus Neural Basis of Learning – Insect Inspired Robots* (Grant No. 01GQ0941).

**Figure 1**

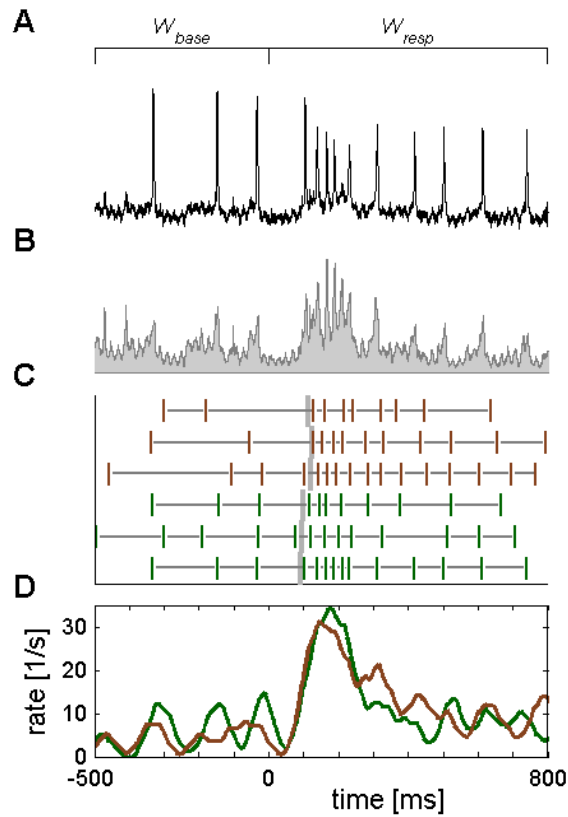

**Fig 1:** Estimation of odor response features. (A) Exemplified odor-response trace of one identified LN. $W_{base}$ indicates the 500ms pre stimulus onset interval which was used to calculate baseline activity. $W_{resp}$ indicates the 800ms interval considered for response analysis. (B) Squared sub-threshold signal from the trace as in A, spikes were removed from the trace. (C) To estimate the mean cell latency (black line), spike trains were first aligned within repeated odor stimulations (red/green) and subsequently across stimuli. Single trial latencies are indicated by vertical gray bars. The CV2 was calculated based on the inter spike intervals (horizontal grey bars). (D) Time-resolved firing rate profiles estimated from repeated and aligned trials to two different odor stimuli (red, green).

**Figure 2**

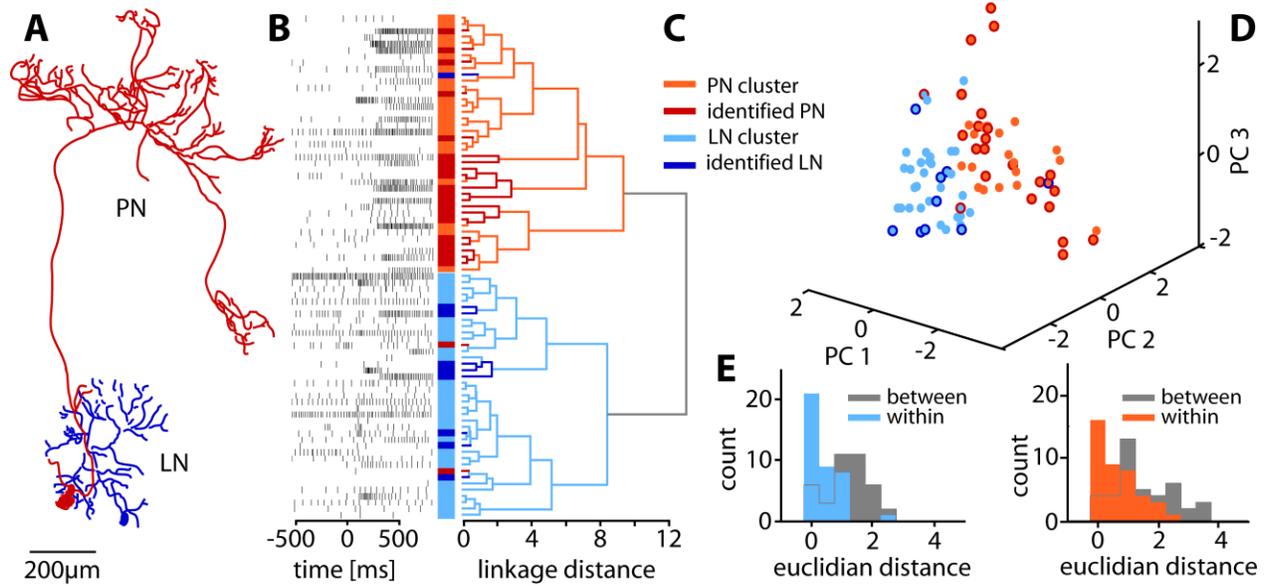

**Fig 2:** Classification of AL neurons based on physiological response features. (A) Morphological reconstructions of one PN (dark red) and one LN (dark blue) contained in the analyzed dataset. (B) Exemplary spike trains (left) randomly selected to illustrate each single neuron's activity. (C) Based on $\Delta R$, L, CV2, FF, and $P_{base}$ identified PNs (dark red) and identified LNs (dark blue) group into a PN dominated cluster (light red) and a PN dominated cluster (light blue). (D) Scatter plot of PN and LN cluster in three dimensional PC space. Data points corresponding to morphologically identified PNs/LNs are marked in dark red (PNs) and dark blue (LNs), respectively. (E) Distribution of distance from individual data points to cluster centers within and between clusters.

**Figure 3**

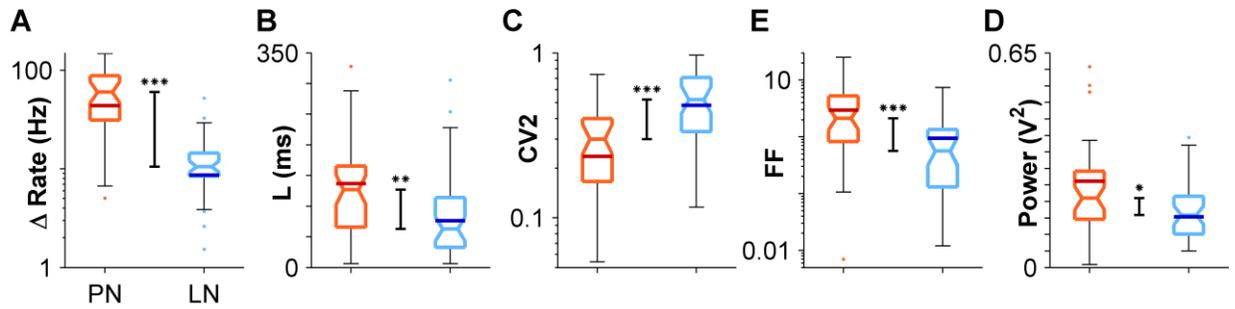

**Fig 3:** LN and PN clusters differ significantly in all five response features. Box-plots illustrating the features distributions for cells in the PN (light red) and in the LN cluster (light blue). Dark red and dark blue horizontal bars indicate medians of the morphologically identified neurons. Differences between populations were significant in all cases (Wilcoxon rank sum test; *p = 0. 05, ** p = 0. 01,*** p = 0. 001).